\documentclass[prl,twocolumn,superscriptaddress,showpacs,floatfix]{revtex4}
\usepackage[dvips]{graphicx}

\begin{document}
%
\title{{\cal Ab initio} coupled-cluster study of $^{16}$O}
\author{M.~W{\l}och}
\affiliation{Department of Chemistry,
Michigan State University, East Lansing, MI 48824, USA}

\author{D.~J.~Dean}
\affiliation{Physics Division, Oak Ridge National Laboratory,
P.O. Box 2008, Oak Ridge, TN 37831, USA}

\author{J.~R.~Gour}
\affiliation{Department of Chemistry,
Michigan State University, East Lansing, MI 48824, USA}

\author{M.~Hjorth-Jensen}
\affiliation{Department of Physics and Center of Mathematics for Applications,
University of Oslo, N-0316 Oslo, Norway}

\author{K.~Kowalski}
\affiliation{Department of Chemistry,
Michigan State University, East Lansing, MI 48824, USA}

\author{T.~Papenbrock}
\affiliation{Physics Division, Oak Ridge National Laboratory,
P.O. Box 2008, Oak Ridge, TN 37831, USA}
\affiliation{Department of Physics and Astronomy, University of
Tennessee, Knoxville, TN 37996, USA}

\author{P.~Piecuch}
\affiliation{Department of Chemistry,
Michigan State University, East Lansing, MI 48824, USA}
\date{\today}
\begin{abstract}
We report converged results for the ground and excited states
and matter density of $^{16}$O using realistic two-body
nucleon-nucleon interactions and
coupled-cluster methods and formalism developed in quantum chemistry.
Most of the binding is obtained with the
coupled-cluster singles and doubles approach. Additional binding
due to three-body clusters (triples) is minimal.
The coupled-cluster method with singles and doubles
provides a good description of the matter density, charge radius,
charge form factor,
and excited states of a 1-particle-1-hole nature, but it
cannot describe the first excited $0^{+}$ state. Incorporation
of triples has no effect on the latter finding.
\end{abstract}

\maketitle

One of the most important problems in nuclear physics
is to understand
how nuclear properties arise from the underlying nucleon-nucleon
interactions. Recent progress using
Monte Carlo \cite{pieper02} and diagonalization \cite{navratil02}
techniques produced converged results for nuclei with up
to $A=12$ active particles,  yielding a much-improved
understanding of nuclear forces in light systems.
One also must explore alternative methods
that would not suffer from the exponential growth
of the configuration space, enabling accurate {\it ab initio}
calculations for medium-size nuclei.
Coupled-cluster theory
\cite{coestercizek} is a promising
candidate for such developments since it 
provides an accurate description
of many-particle correlations at relatively low cost,
as has been demonstrated in numerous chemistry
applications \cite{chem_rev,piecuch02}.
Recently, Mihaila and
Heisenberg performed coupled-cluster calculations for the binding
energy and the electron scattering form factor of $^{16}$O
using bare interactions \cite{bogdan}. In previous work \cite{kowalski04}, we took another
route and used
quantum chemical coupled-cluster methods and the
renormalized Hamiltonian to compute ground
and excited states of $^4$He and ground-state energies of $^{16}$O in a
small model space consisting of 4 major oscillator shells,
demonstrating promising results when compared with
exact shell-model diagonalization.

In this Letter we report, for the first time, converged coupled-cluster
calculations for ground- and excited-state energies and other
properties of $^{16}$O using modern
nucleon-nucleon interactions derived from
effective-field theory \cite{eft}.  Our ground-state calculations involving
one- and two-body components of the cluster operator are performed in
up to 8 major
oscillator shells (480 uncoupled single-particle basis states), while
the corrections due to three-body clusters and
computations of excited states and nuclear properties
involve up to 7 major oscillator shells (336
single-particle states). The significant progress in going from
model calculations
using 80 single-particle states \cite{kowalski04}
to large-scale calculations involving 16 correlated
nucleons and almost 500 single-particle states has
been possible thanks to the development of
general-purpose coupled-cluster computer
programs for nuclear structure,
using diagram factorization techniques adopted by quantum chemists.
We pay particular attention to three aspects of the calculations:
(i) the convergence of the ground-state
energy with respect to the size of the model space and the role
of higher--than--two-body clusters in such studies,
(ii) the ability of coupled-cluster methods to
describe excited states, and (iii) the performance of
coupled-cluster methods in studies of nuclear radii,
matter density, charge form factor, and occupation numbers.
We have not yet included the three-nucleon interaction
that should eventually be considered \cite{pieper02,navratil02}.
However,
our calculations represent a dramatic step forward in nuclear
many-body computations due to the enormous oscillator space we probe
through application of computationally efficient coupled-cluster methods.
They teach us about the
nucleon correlations and the magnitude of the (missing)
three-body forces.

We use two variants of effective-field-theory-inspired Hamiltonians,
Idaho-A and N3LO \cite{entem2002}.  The Idaho-A potential was derived
with up to chiral-order three diagrams while N3LO includes
chiral-order four diagrams, and charge-symmetry and
charge-independence breaking terms. We also include the Coulomb
interaction with the N3LO calculations.  Since very slow convergence
with the number of single-particle basis states was obtained using
bare interactions \cite{bogdan}, we renormalize the bare Hamiltonian
using a no-core G-matrix approach \cite{dean04} which obtains a
starting-energy dependence $\tilde{\omega}$ in the two-body matrix
elements $G(\tilde{\omega})$.  We use the Bethe-Brandow-Petschek
\cite{bbp63} theorem to alleviate much of the starting-energy
dependence (see \cite{dean04} for details). The dependence upon
the starting energy is weak for $^{16}$O, particularly for the
matrix elements below the Fermi surface \cite{ellis94}.  The effective
Hamiltonian for coupled-cluster calculations is
$H^{\prime}=t+G(\tilde{\omega})$, where $t$ is the kinetic energy.  We
correct $H^{\prime}$ for center-of-mass contaminations using the expression
$H=H^{\prime}+\beta_{\rm c.m.}H_{\rm c.m.}$.  We choose
$\beta_{\rm c.m.}$ such that the expectation value of the
center-of-mass Hamiltonian $H_{\rm c.m.}$ is 0.0 MeV.  We note that
intrinsic excitation energies are virtually independent of $\beta_{\rm
c.m.}$ while the unphysical, center-of-mass contaminated states show a
sharp, nearly linear dependence of excitation energies on $\beta_{\rm
c.m.}$.  This allows us to separate intrinsic and center-of-mass
contaminated states.

Once the one- and two-body matrix elements of the
center-of-mass-corrected effective Hamiltonian
are constructed, we solve the $A$-body problem
using quantum chemical coupled-cluster techniques.
In the ground-state calculations,
we use the CCSD (``Coupled-Cluster Singles and
Doubles'') approach \cite{purvis82}, to describe
correlation effects due to one-
and two-body clusters, and the
CR-CCSD(T) (``Completely Renormalized
CCSD(T)'') method
\cite{kowalski2004},
to correct the CCSD energies for the effects of three-body
clusters (``Triples'').
In the excited-state and property calculations, we use
the equation-of-motion (EOM) CCSD method \cite{stanton93}
(equivalent to the linear response
CCSD approach
\cite{monkhorst77}).
We also
correct the energies of excited states obtained with
EOMCCSD for the effects of triples using the
CR-EOMCCSD(T) approach \cite{kowalski2004}.
The details of the above methods can be found elsewhere
\cite{purvis82,stanton93,kowalski2004}.  Here, we only mention that
the CCSD method is obtained by truncating the many-body expansion for
the cluster operator $T$ in the exponential ansatz exploited in
coupled-cluster theory, $\mid\Psi_{0}\rangle =
\exp(T)\mid\Phi\rangle$, where $\mid\Psi_{0}\rangle$ is the correlated
ground-state wave function and $\mid\Phi\rangle$ is the reference
determinant.
The truncated cluster operator used in the CCSD calculations has the
form $T = T_{1} + T_{2}$, where $T_1=\sum_{i,a} t_a^i a^\dagger_{a} a_{i}$ and
$T_2= \frac{1}{4} \sum_{ij,ab} t_{ab}^{ij} a^\dagger_{a} a^\dagger_{b} 
a_{j} a_{i}$
are the singly and doubly excited clusters and $i,j,\ldots$
($a,b,\ldots$) label the single-particle states occupied (unoccupied)
in $|\Phi\rangle$. 
We determine the singly and
doubly excited cluster amplitudes $t_a^i$ and $t_{ab}^{ij}$
by solving the nonlinear system of algebraic equations, $\langle
\Phi_{i}^{a} | \bar{H}|\Phi\rangle = 0$, $\langle \Phi_{ij}^{ab} |
\bar{H}|\Phi\rangle = 0$, where $\bar{H} = \exp(-T) \, H \exp(T)$ and
$|\Phi_{i}^{a}\rangle$ and $|\Phi_{ij}^{ab}\rangle$ are the singly and
doubly excited determinants, respectively, relative to
$|\Phi\rangle$.  
We calculate the ground-state energy $E_0$ as 
$\langle\Phi\mid\bar{H}\mid\Phi\rangle$.
We diagonalize the similarity-transformed Hamiltonian
$\bar{H}$ in the relatively small space of singly and doubly excited
determinants $|\Phi_{i}^{a}\rangle$ and $|\Phi_{ij}^{ab}\rangle$ to
obtain the excited-state wave functions $|\Psi_{\mu}\rangle$ and
energies $E_{\mu}$. The right eigenstates of $\bar{H}$, $R^{(\mu)} |
\Phi\rangle $, where $R^{(\mu)} = R_{0}+ R_{1} + R_{2}$ is a sum of
the relevant reference ($R_{0}$), one-body ($R_{1}$), and two-body
($R_{2}$) components define the excited-state ``ket'' wave functions
$|\Psi_{\mu}\rangle=R^{(\mu)} \exp(T) |\Phi\rangle$, whereas the left
eigenstates $\langle \Phi | L^{(\mu)}$ define the ``bra'' wave
functions $\langle \tilde{\Psi}_{\mu}| = \langle \Phi | L^{(\mu)}
\exp(-T)$.  Here, each $n$-body component of $R^{(\mu)}$ with $n>0$ is
a particle-hole {\it excitation} operator similar to $T_{n}$, whereas
$L^{(\mu)}$ is a hole-particle {\it deexcitation} operator, so that
$L_1=\sum_{i,a} l_i^a a^\dagger_{i} a_{a}$ and $L_2= \frac{1}{4} \sum_{ij,ab}
l_{ij}^{ab} a^\dagger_{i} a^\dagger_{j} a_{b} a_{a}$.  The right and 
left eigenstates
of $\bar{H}$ form a biorthonormal set, $\langle \Phi | L^{(\mu)} \,
R^{(\nu)} | \Phi\rangle = \delta_{\mu\nu}$.  If the only purpose of
the calculation is to obtain excitation energies, the left eigenstates
$\langle \Phi | L^{(\mu)}$ are not needed.  However, for properties
other than energy, both right and left eigenstates of $\bar{H}$ are
important.  In particular, we calculate the one-body reduced density
matrix $\rho_{\alpha \beta}$ in quantum state $|\Psi_{\mu} \rangle$ as
follows:
\begin{equation}
\rho_{\alpha \beta} =
\langle \Phi | L^{(\mu)} \,\left[\exp(-T) \, a^\dagger_{\alpha}
a_{\beta} \exp(T)\right]
\, R^{(\mu)} 
|\Phi \rangle\;.
\label{prop1}
\end{equation}
In the CCSD ground-state ($\mu=0$) case, we have
$T=T_{1}+T_{2}$, $R^{(0)} = 1$, and $L^{(0)} = 1 + \Lambda_{1} +
\Lambda_{2}$, where the one- and two-body deexcitation operators
$\Lambda_{1}$ and $\Lambda_{2}$ are determined by solving the CCSD
left eigenvalue problem, obtained by right-projecting the equation
$\langle\Phi\mid(1+\Lambda) \bar{H} = E_{0} \langle \Phi |
(1+\Lambda)$, with $E_{0}$ representing the CCSD energy and $\Lambda =
\Lambda_{1} + \Lambda_{2}$ on the singly and doubly excited
determinants.  Thus far, we have focused on the CCSD
and EOMCCSD methods which
use
inexpensive computational steps that scale as
$n_{o}^{2} n_{u}^{4}$, where $n_{o}$ ($n_{u}$) is the number of
occupied (unoccupied) single-particle states.
While the full inclusion of triply excited clusters 
is possible, the resulting
methods are expensive and scale as $n_{o}^{3} n_{u}^{5}$.
Thus,
we estimate the effects of $T_{3}$ and $R_{3}$
on ground- and excited-state energies by adding
the corrections
to the CCSD/EOMCCSD energies,
which only require $n_{o}^{3} n_{u}^{4}$ noniterative steps.
These corrections, due to $T_{3}$ and $R_{3}$,
define the
CR-CCSD(T) and CR-EOMCCSD(T) approaches \cite{piecuch02,kowalski2004}.
In this study, we use variant ``c'' of the
CR-CCSD(T) and CR-EOMCCSD(T) approaches \cite{kowalski04}.

\begin{figure}[hbpt]
\includegraphics[ scale=0.25, angle=270]{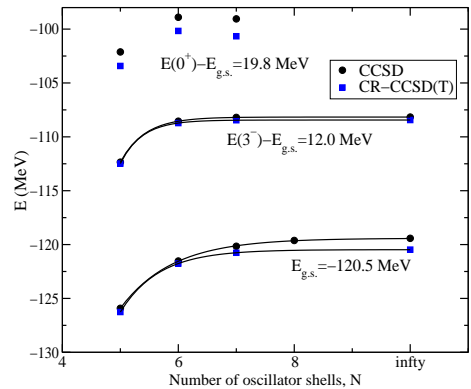}
\caption{The energies of the ground-state (g.s.) and first-excited 
$3^-$ and $0^+$ states as functions of the number
of oscillator shells $N$ obtained with coupled-cluster methods and the
Idaho-A interaction.
}
\label{fig1}
\end{figure}

We turn to a discussion of our $^{16}$O results.  We choose the
oscillator energy $\hbar\omega$ for our basis states to minimize the
CCSD energy. For the $N=7$ and $N=8$ oscillator shell runs,
$\hbar\omega=11$~MeV, and the results are nearly 
independent of $\hbar\omega$ \cite{dean04}.  
Shown in Fig.~\ref{fig1} are our CCSD/EOMCCSD and
CR-CCSD(T)/CR-EOMCCSD(T) ground- and excited-state energies as a
function of $N$. The symbols in Fig.~\ref{fig1} represent our
calculations while the lines represent a fit of the form
$E(N)=E_\infty+a\exp\left(-b N\right)$, where the extrapolated energy
$E_\infty$ and $a$ and $b$ are parameters for the fit. We also show in
Fig.~\ref{fig1} our calculations for the first excited $3^-$ state and
the position of the lowest calculated $0^+$ excited state. We now discuss 
these results.

{\it Triples correction to the CCSD ground-state energy.}  The small model space calculation \cite{kowalski04} indicated that the triples corrections to
the ground-state CCSD energies
are small. We extended
these calculations from 4 to 8 major oscillator shells for CCSD calculations
and to 7 major oscillator shells for CR-CCSD(T) calculations, as shown in
Fig.~\ref{fig1}.  We find that the extrapolated CCSD energy is
$-119.4$~MeV for Idaho-A.  For the $N=7$ Idaho-A calculation, the
difference between the CCSD and CR-CCSD(T) result is 0.6~MeV, while
the extrapolated values differ by only 1.1~MeV; our extrapolated
CR-CCSD(T) energy is $-120.5$~MeV.  The Coulomb interaction adds
to the binding 11.2~MeV, so that our estimated Idaho-A ground state
energy is $-109.3$~MeV (compared to an experimental value of
$-128$~MeV). Our $N=7$ ($N=8$) N3LO CCSD and $N=7$ 
CR-CCSD(T) energies, which
include the Coulomb interaction, are $-112.4$ ($-111.2$) and $-112.8$~MeV,
respectively.  Thus, the two-body interactions underbind $^{16}$O by
approximately 1~MeV per particle, pointing to the need 
for three-body forces.  For the Idaho-A and N3LO interactions and the
$^{16}$O nucleus, we conclude that connected $T_{3}$ clusters are
indeed small, contributing less than 1\% to the ground-state
energy. This is an important finding, since it implies that
essentially all correlations in a closed-shell nucleus resulting from
two-body nucleon-nucleon interactions can be captured by the
relatively inexpensive CCSD approach. Another important finding is a
rapid convergence of the CCSD and CR-CCSD(T) energies with the number
of oscillator shells owing to the use of the renormalized form of the
Hamiltonian. For example, the difference between the $N=8$ and $N=7$
CCSD/Idaho-A energies is 0.5 MeV (see Fig.~\ref{fig1}).

{\it Calculations of the first excited $3^-$ state.}  The first
excited $3^-$ state in $^{16}$O is thought to be principally a
one-particle-one-hole ($1p\mbox{-}1h$) state \cite{wb92}. The experience 
of quantum chemistry
is that the EOMCCSD and CR-EOMCCSD(T) methods describe such states
well, provided that the three-body interactions in the Hamiltonian can
be ignored.  The largest $R_1$ amplitudes obtained in the EOMCCSD
calculations indicate that the dominant $1p\mbox{-}1h$ excitations are
from the $0p_{1/2}$ orbital to the $0d_{5/2}$ orbital. The
$2p\mbox{-}2h$ excitations in the EOMCCSD wave function, defined as
$R_2 + R_{1}T_{1}+R_{0}(T_{2}+T_{1}^2/2)$ ($R_{0}=0$ in this case), are much
smaller than the $R_1$ amplitudes, and the CR-EOMCCSD(T) calculation
hardly changes the total energy of the state, which indicates that
this state has indeed a $1p\mbox{-}1h$ nature. Our extrapolated
Idaho-A results indicate that the $3^-$ state lies at $-108.2$ and
$-108.4$~MeV in the EOMCCSD and CR-EOMCCSD(T) calculations,
respectively. The CR-EOMCCSD(T) method yields an excitation energy of
12.0~MeV for this state which experimentally lies at 6.12~MeV. N3LO
yields similar results. Based on the $1p\mbox{-}1h$ structure of the
state, we conclude that Idaho-A and N3LO do not yield an excitation
energy for the $3^-$ state which is commensurate with experiment.
These results agree with recent no-core shell-model calculations with
similar two-body Hamiltonians \cite{petrprivate2004}.  
The $3^-$ state is expected to be built on $1p\mbox{-}1h$ 
excitations which depend
on the single-particle splittings. These splittings will be affected
by three-body forces not included in our Hamiltonian, thus affecting the
energy of the $3^-$ state. 
Whether other mechanisms than three-body
forces can provide an additional binding of 6 MeV needs further research.
Our results are
converged at the coupled-cluster level employing the Idaho-A and
N3LO two-body interactions, so it is likely that the discrepancy
between theory and experiment resides in the Hamiltonian, not in the
correlation effects which EOMCCSD and CR-EOMCCSD(T) describe very well
if three-body forces play no role and if the state has a
$1p\mbox{-}1h$ nature.

{\it Calculation of the first excited $0^{+}$ state}.  This state
(experimentally at 6.05~MeV), believed to have a $4p$-$4h$ character,
cannot be described by EOMCCSD or CR-EOMCCSD(T). This is confirmed 
by our calculations as we see large
differences between the EOMCCSD or CR-EOMCCSD(T) results and experiment (see
Fig.~\ref{fig1}). 
One would need to include $4p\mbox{-}4h$ operators ($T_{4}$ and $R_{4}$)
to improve coupled-cluster results.

Although we concentrated on the lowest energy $J=3^-$ and $J=0^+$
excited states and the role of three-body clusters on these, we also
performed preliminary calculations for other negative parity
states. The quartet of negative parity states starting with the
$J=3^-$ state, and including the $J=1^-,2^-$ and $0^-$ states, are all
believed to have a similar $1p\mbox{-}1h$ character \cite{wb92}. The
EOMCCSD calculation with 5 major oscillator shells and Idaho-A
confirms the existence of this quartet, giving excitation energies of
13.57, 15.37, 17.07, and 17.15 MeV for the $J=3^-$, $1^-$, $2^-$, and
$0^-$ states, respectively.  While these states are all a few MeV above the
experimental values, their ordering predicted by EOMCCSD is correct.

\begin{figure}[hbpt]
\includegraphics[scale=0.3]{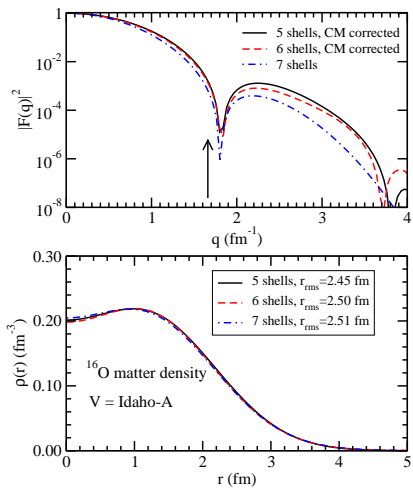}
\caption{Top panel: The charge form factor computed from the CCSD
density matrix.  Bottom panel: the matter density in $^{16}$O.  The
results obtained with the Idaho-A interaction.}
\label{fig2}
\end{figure}

{\it Calculation of the one-body density.}  We use Eq.~(\ref{prop1}),
where $\mu=0$, to calculate the ground-state density for $^{16}$O. 
We show
the resulting radial density, $\rho(r)$,
in Fig.~\ref{fig2}. The root-mean-square (rms)
radius is found through an integration $r_{\rm rms}^2 = \int
r^4\rho(r)dr / \int r^2 \rho(r) dr$.  To obtain a charge radius, we correct this value for
the finite size of the nucleons, which experimentally are
$r_p^2=0.743$~fm$^2$ and $r_n^2=0.115$~fm$^2$, and for the $0s$
center-of-mass motion, for which we use $\langle\Psi_{0}\mid {\bf
R}\mid \Psi_{0}\rangle=\frac{62.2071}{A\hbar\omega}$~fm$^2$. Our rms
charge radii for $^{16}$O for 5, 6, and 7 oscillator shells are
2.45~fm, 2.50~fm, and 2.51~fm, respectively when the Idaho-A
potential is used (N3LO gives similar values).  The experimental
charge radius is 2.73$\pm$0.025~fm. We also calculated the occupation
probability for the natural orbitals. Experimental data from
quasi-elastic proton knockout \cite{leuschner94} yields $2.17\pm
0.12$\% for the $0d_{5/2}$ occupation and $1.78\pm 0.36$\% for the
$1s_{1/2}$ occupation.  We obtain 3.2\% and 2.3\% respectively, using
Idaho-A in the $N=7$ model space. For N3LO in the $N=7$ model space,
we obtain 3.8\% and 2.6\%, respectively.
For the calculation of the nuclear charge form factor, we follow
\cite{bogdan00}. In this approach, the form factor includes
contributions from the two-body reduced density matrix due to
center-of-mass corrections. We computed the one-body density
contributions within the framework of CCSD theory using Eq.~(\ref{prop1}). The
contributions of the two-body density matrix were computed within the
shell-model like description as $\rho_{\alpha\beta\gamma\delta} =
\langle\Psi_{0}\mid a^\dagger_{\alpha} a^\dagger_{\beta} 
a_{\delta} a_{\gamma} \mid
\Psi_{0} \rangle/\langle\Psi_{0}|\Psi_{0} \rangle$, where we
approximated $\mid\Psi_{0}\rangle$ by
$\left(1+C_1+C_2\right)\mid\Phi\rangle$, with $C_{1}=T_{1}$ and $C_{2}
=T_{2}+\frac{1}{2} T_1^2$ defining the $1p\mbox{-}1h$ and
$2p\mbox{-}2h$ components of the CCSD wave function.
The upper part of Fig.~\ref{fig2} shows the charge form factor for
different model spaces. The 5-shell and
6-shell results include the center-of-mass corrections and exhibit a
second zero. 
Compared to the experimental value (the arrow in Fig.~\ref{fig2}), 
the first zero of the form-factor is reasonable, although slightly 
too large; this is consistent with an underestimated value 
of the theoretical charge radius.

In summary, the $^{16}$O ground state is converged with respect to the
model space size and is accurately described within the basic CCSD
approximation, with three-body clusters contributing less than 1\% of
the binding energy. We attribute the 1~MeV per particle difference
between the coupled-cluster and experimental binding energies to
three-body forces. We obtained a correct description of the quartet of
low-lying negative parity $1p\mbox{-}1h$ excited states, although there is a
6-MeV difference between the converged coupled-cluster results and
experiment for the lowest $J=3^-$ state, which is, quite likely, due to
an inadequate description of the relevant nuclear forces by the
Hamiltonian. We were unable to accurately describe the lowest $J=0^+$
excited state due to connected $4p\mbox{-}4h$ correlations missing in
coupled-cluster approximations employed in this study.  The CCSD
method provides reasonable results for the nuclear matter density,
charge radius, and charge form factor. The
use of the renormalized Hamiltonian guarantees fast convergence of the
results with the number of oscillator shells. All of this 
makes low-cost coupled-cluster methods a promising 
alternative to traditional shell-model diagonalization techniques.

Research supported by 
the U.S. Department of Energy (Oak Ridge National Laboratory,
University of Tennessee, Michigan State University),
the National Science Foundation (Michigan State University), and
the Research Council of Norway (University of Oslo).

\end{document}